\documentclass[sn-basic]{sn-jnl}

\theoremstyle{thmstyleone}%
\theoremstyle{thmstyletwo}%
\usepackage{lineno}
\theoremstyle{thmstylethree}%
\usepackage{appendix}
\usepackage[english]{babel}
\usepackage{amsmath, amsfonts, amssymb, mathrsfs}
\usepackage{subfigure}

\begin{document}

\title [\hspace{1 pt}]{ Environmental Toxicity Influences Disease Spread In Consumer Population
}


\author[1]{\fnm{Arnab} \sur{Chattopadhyay}} 

\author*[1,2]{\fnm{Swarnendu} \sur{Banerjee}} \email{swarnendubanerjee92@gmail.com}

\author[1]{\fnm{Amit} \sur{Samadder}} 

\author*[1]{\fnm{Sabyasachi} \sur{Bhattacharya}} \email{sabyasachi@isical.ac.in}

\affil[1]{\orgdiv{Agricultural and Ecological Research Unit}, \orgname{Indian Statistical Institute}, \orgaddress{\street{203, B. T. Road}, \city{Kolkata}, \postcode{700108}, \state{West Bengal}, \country{India}}}

\affil[2]{\orgdiv{Copernicus Institute of Sustainable Development}, \orgname{ Utrecht University}, \orgaddress{\street{ PO Box 80115, 3508
TC Utrecht, the Netherlands}}}


\abstract{ The study of infectious disease has been of interest to ecologists since long. The initiation of epidemic and the long term disease dynamics are largely influenced by the nature of the underlying consumer (host)-resource dynamics. Ecological traits of such systems may be often modulated by toxins released in the environment due to ongoing anthropogenic activities. This, in addition to toxin-mediated alteration of epidemiological traits, has a significant impact on disease progression in ecosystems which is quite less studied. In order to address this, we consider a mathematical model of disease transmission in consumer population where multiple traits are affected by environmental toxins. Long term dynamics show that the level of environmental toxin determines disease persistence, and increasing toxin may even eradicate the disease in certain circumstances. Furthermore, our results demonstrate bistability between different ecosystem states and the possibility of an abrupt transition from disease-free coexistence to disease-induced extinction of consumers. Overall the results from this study will help us gain fundamental insights into disease propagation in natural ecosystems in the face of present anthropogenic changes.}


\maketitle

\section{Introduction}
\label{introduction}

Environmental pollution and disease outbreaks are global threat to ecosystems in the present era of the Anthropocene \citep{lafferty2004diseases, van2009emerging}. Toxins released in the environment due to anthropogenic factors have long-drawn consequences on the ecosystem health \citep{huang2013model, huang2015impact, garay2013more, banerjee2021chemical}. The emergence of disease outbreaks in many ecosystems is also of utmost concern for ecologists \citep{lafferty1999environmental, lafferty2003should,  lafferty2004diseases, khan1990parasitism, van2009emerging}. It is well known that environmental toxin can influence the impact of disease spread on ecosystem but the manner of it is not properly understood. Toxin can reduce the host immunity thus making them more susceptible which increases disease prevalence \citep{khan1990parasitism, beck2000host}. Conversely, toxin may restrict movement and increase mortality of the infected host thus having a negative effect on disease spread. These synergistic and antagonistic effect makes it difficult to predict the impact of environmental toxin on long term disease dynamics in ecosystem thus highlighting the need for studies combining the two.

While toxin can affect disease transmission rate in multiple ways, it can also impact host-resource interaction which in turn can have significant impact on disease dynamics \citep{huang2015impact}. In fact, the host-resource dynamics has received a lot of attention in the recent years in regard to disease progression in ecosystems \citep{hurtado2014infectious,hilker2008disease}. In this context, it is important to note that even life history traits like growth rate and mortality of both resource and the host and the ingestion rate of the later can be influenced by environmental toxin. This affects abundance of the host population and thus disease prevalence. Majority of theoretical studies which tried to address the impact of chemical pollution on disease however neglected these effects \citep{sinha2010two, liu2012dynamics, chauhan2015effect, wang2004persistence}. Although a recent study by \cite{banerjee2019effect} on a \emph{Daphnia}-algae-fungus system took into consideration some of these aspects, it has few shortcomings. These are mainly based on the fact that the toxic effect considered in this study are system specific which may not be reasonable when extrapolated to other contaminated systems in the environment. 

For instance, \cite{banerjee2019effect} accounted for reduced disease transmission due to behavioural change of the host at toxic concentration of the pollutant. However, high environmental toxin may also lead to increased susceptibility of the host to disease \citep{coors2008synergistic, beck2000host}. This may be a result of toxin-induced immunity suppression which is ubiquitous in many epidemic scenarios \citep{coors2008synergistic} and well demonstrated in plenty of experimental studies \citep{de1994impairment, de1996impaired, ross1996contaminant, bogomolni2016vitro}. Especially, aquatic species and more specifically many marine mammals are known to be vulnerable to immuno-toxic contaminants \citep{ross2000marine, ross2002role, de1994impairment, de1996impaired, ross1996contaminant, bogomolni2016vitro}. For instance, harbor seals would be less immune if they fed on fish from the more polluted Baltic Sea and less susceptible in case their predation is associated with the less polluted Atlantic sea \citep{ross1996contaminant, de1994impairment}.  Furthermore, it has been pointed out earlier that environmental toxin should also have an effect on the carrying capacity as well \citep{freedman1991models}, which was not taken into account by \cite{banerjee2019effect}. 

To address these gaps, we model disease progression in consumer population using the approach employed by earlier studies like \cite{huang2013model,huang2015impact, thieme2003princeton}.  This implied the use of Beverton-Holt growth rate instead of conventional logistic formulation. Additionally, we consider toxin dependent increase in transmission and attempt to answer two interrelated questions: (1) How does environmental toxin influence the progression of disease? (2) How do the contamination and disease jointly shape the community composition of an ecosystem? First, we describe our model and toxin-mediated response functions in section \ref{methods}. Also, we make our model parameters dimensionless and use a quasi-steady state approximation in this section to reduce model complexity. In section \ref{results}, we investigate the possible asymptotic states of our system with the help of bifurcation diagrams. Finally, our paper is summarized with a brief discussion in section \ref{discussion}.

\section{Methods} \label{methods}

\subsection{Model description}

We consider a consumer-resource model, where the consumer population is affected by an infectious disease \citep{hilker2008disease}. Furthermore, we assume the ecological and epidemiological traits of the system are altered by environmental toxins. Let $x(t)$ and $y(t)$ be the concentration of the resource and the consumer biomass respectively. The consumer (host) population can be further segregated into susceptible, $S(t)$, and infected, $I(t)$, classes such that $S(t)+I(t)= y(t)$. It must be noted that, we have used the term `consumer' and `host' interchangeably through out the paper. Let $u(t)$ and $v(t)$ be the toxin body burden of the resource and consumer species respectively, which is defined as the ratio of the total toxin in a population to the total biomass concentration \citep{huang2015impact}. Then the disease dynamics under the influence of environmental toxin can be described as below:

\begin{subequations}
\label{eq1}

    \begin{equation}
       \frac{dx}{dt}= (\beta(x,u) -\mu_{1}(u))x-\frac{ax}{H+x}y \\
    \end{equation}
    
    \begin{equation}
       \frac{dS}{dt}= e(v)\frac{ax}{H+x}y-\lambda(v) \frac{ SI}{y}-\mu_{2}(v)S \\
    \end{equation}
    
    \begin{equation}
      \frac{dI}{dt}=\lambda(v) \frac{ SI}{y}- \mu_{2}(v)I-\eta I  \\
      \end{equation}
    
    \begin{equation}
        \frac{dy}{dt}=\frac{d(S+I)}{dt}= e(v)\frac{ax}{H+x}y-\mu_{2}(v)y-\eta I \\
    \end{equation}

\end{subequations}

The first term on the right hand side of the equation \ref{eq1}(a) represents the net growth of the resource species in the absence of consumers, which is taken to be Beverton-Holt type (see \cite{thieme2003princeton} for derivation of this term). Here, the reproduction and growth rate is represented by $\beta(x,u)=\displaystyle \frac{\alpha_{1} }{1+ \alpha_{3}x}  b(u)$, where $b(u)$ is the effect of toxin on the resource's growth rate. The death rate is denoted by $\mu_{1}(u)$ which depends on the toxin body-burden of the resource, $u$. The second term is the biomass loss due to the predation by consumer and the functional response is taken to be Holling type-II, where $a$ is the maximum feeding rate and $H$ is the half-saturation constant. In equations \ref{eq1}(b-c), $e(v)$ is the food conversion efficiency, $\lambda(v)$ denotes the rate of disease transmission and $\mu_{2}(v)$ is the natural mortality rate of the consumer. All these parameters are assumed to be dependent on toxin body burden of the consumer, $v$. Infected consumers have an additional disease-induced death term (virulence), $\eta$. In our model, disease transmission is assumed to be frequency-dependent which implies that the per-capita force of infection increases with disease prevalence, $\displaystyle \frac{I}{y}$ \citep{de1998modelling, hilker2008disease}. 

For simplicity, we assume that the transmission rate is the only epidemiological trait which is toxin dependent. Also, we do not consider any recovery from the disease. For readers' convenience, we used largely the same notations as in \cite{hilker2008disease} and \cite{huang2015impact} for parameters and state variables in this paper. 

\begin{figure}[h!]
    \begin{center}
        {\includegraphics[width=1\textwidth]{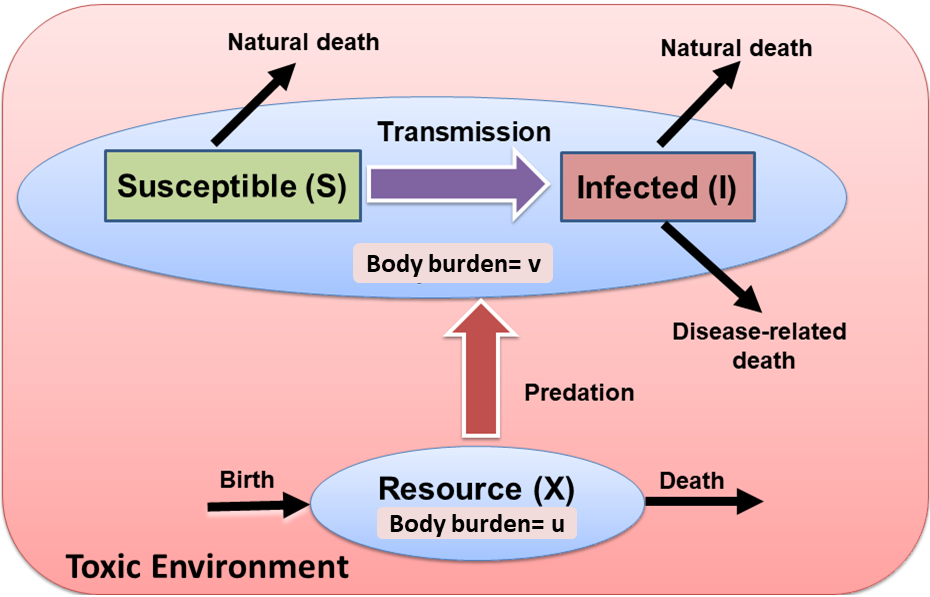}}
    \end{center}
    \caption{ Schematic representation of our model. Resource growth is regulated by birth, death, and predation by the consumer. The consumer is divided into susceptible and infected classes. The system is under the influence of environmental toxin and the body  burden of the resource and consumer are $u$ and $v$ respectively.}
    \label{schematic}
\end{figure}

From the above equation \ref{eq1}, the rate of change of disease prevalence, $i(t) =\displaystyle \frac{I(t)}{y(t)}$, can be expressed as follows (see Appendix \ref{appendixA} for the derivation): 

\begin{equation}
\label{eq2}
    \frac{di}{dt}= \lambda (v) i(1-i)-e(v)\frac{ax}{H+x}i-\eta i(1-i)\\
\end{equation}

Following \cite{hilker2008disease}, we express the model (equation \ref{eq1}-\ref{eq2}) in terms of $x$, $y$ and $i$, i.e., equation \ref{eq1}(a, d) and \ref{eq2}. This is particularly advantageous not only because it helps to remove the singularity in the disease transmission term when host population is zero, but also allows us to establish the cause of extinction of the host population. When host population becomes extinct due ecological factors, for example, high mortality rate the prevalence becomes zero. This is referred to as ecological extinction throughout the text. On the other hand, when epidemiological factors, for example, disease induced death is the underlying mechanism of host extinction, the prevalence remain strictly positive. In this case, which is referred to as epidemiological extinction henceforth, the positive prevalence signifies that the disease transmission occurs even when the population is very small \citep{de2005mechanisms, hilker2008disease}.

\subsection{Modelling toxin accumulation}

In order to incorporate the effect of toxin on the population dynamics of the interacting species, we must track the time evolution of the amount of the accumulated toxin concentration of the resource and consumer species ($U(t)$ and $V(t)$ respectively). Following \cite{huang2015impact}, their dynamical equations can be written as:

\begin{subequations}
 \label{eq3}

    \begin{equation}
        \frac{dU}{dt}= a_{1}Tx-\sigma_{1}U-\mu_{1}(u)U-\frac{axy}{H+x}u \\
    \end{equation}
    \begin{equation}
       \frac{dV}{dt}= a_{2}Ty-\sigma_{2}V+\frac{axy}{H+x}u-\mu_{2}(v)V- \eta i V     \\
    \end{equation}
\end{subequations}

Here, $T$ is the environmental toxicant concentration, $a_{i}$ and $\sigma_{i}$ ($i=1,2$) are the uptake and depuration coefficients of the toxin for resource and consumers respectively. The concentration of toxin accumulated in both the resource and consumer population are regulated by uptake from the environment and depuration due to metabolism. Additionally, toxin is lost due to natural death of both populations and disease induced death of the consumer. Predation by consumer also leads to transfer of toxin from the resource to itself resulting in biomagnification. 

The body-burden of the resource and consumer population, already defined above, can thus be expressed as $u(t)=\displaystyle \frac{U(t)}{x(t)}$ and $v(t)=\displaystyle \frac{V(t)}{y(t)}$ respectively, the rate of change of which is given below (see Appendix \ref{appendixA}):

\begin{subequations}
\label{eq4}
    \begin{equation}
        \frac{du}{dt}= a_{1}T-\sigma_{1}u-\beta(x,u)u
    \end{equation}
    \begin{equation}
       \frac{dv}{dt}= a_{2}T-\sigma_{2}v+\frac{ax}{H+x}(u-ve(v)) 
    \end{equation}
\end{subequations}

The system can now be fully described using  the state variables, $x$, $y$ and $i$ together with the toxin body burdens $u$ and $v$. So the equations \ref{eq1}(a), \ref{eq1}(d), \ref{eq2}, and \ref{eq4} are the equations of our interest for the remaining part of the paper.  

\subsection{Modeling responses due to the toxin}

For analyzing our model, we describe specific forms of toxin body burden dependence for each of the concerned parameters mentioned in the earlier paragraphs.

\begin{figure}[h!]
    \begin{center}
        \includegraphics[width=1\textwidth]{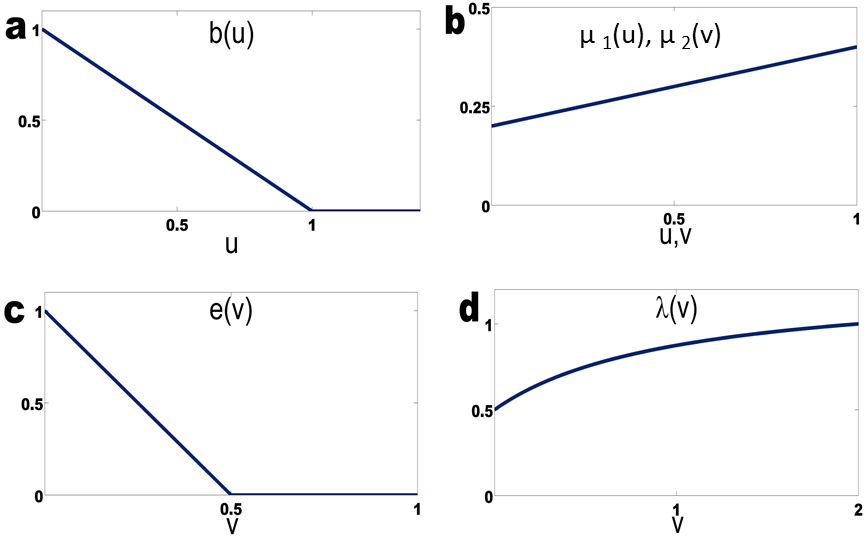}
    \end{center}
    \caption{The responses of toxin body burden on parameters: (a) maximum reproduction rate, (b) mortality of the resource and consumer, (c) conversion efficiency of the consumer, (d) disease transmission rate.}
    \label{responses}
\end{figure}

\begin{table}[h]
\label{table}
 \tiny
 \caption{Variables and parameters description}
 \vspace{0.5 cm}
 \label{par_table}       
 \begin{tabular}{lllll}
 \hline\noalign{\smallskip}
 
Symbols                    & Unit             & Description                                                \\
\noalign{\smallskip}\hline\noalign{\smallskip}
\\
Variables
\\
$x$   & $g/L$  & resource density \\

$y$   & $g/L$ & consumer density \\

$S$   & $g/L$ & susceptible consumer density \\

$I$   & $g/L$ & infected consumer density \\

$U$  & $\mu g/L$  & concentration of the toxin in the resource \\

$V$  & $\mu g/L$  &  concentration of the toxin in the consumer \\ 

$u$  & $\mu g/g$  & body burden of the resource \\

$v$  & $\mu g/g$  & body burden of the consumer \\
\\
Parameters
\\
$\alpha_{1}$ & $day^{-1}$ & maximum reproduction rate of the resource \\ 

$\alpha_{2}$ & $g/\mu g$ & effect of toxin on the growth of resource\\

$\alpha_{3}$ & $L/g$ & crowding effect of resource \\

$k_{1}$ & $g/\mu g/day$ & effect coefficient of the toxin on the resource mortality \\

$p$ & $day^{-1}$ & natural mortality of resource \\

$a$ & $day^{-1}$ & per-capita feeding rate\\

$H$ & $g/L$ & half saturation constant\\
\\

$\beta_{1}$ & {-} & reproduction efficiency of consumer\\

$\beta_{2}$ & $g/\mu g$ & effect of toxin on the reproduction of consumer\\

$m$ & {-} & effect coefficient of the toxin on the transmission of the disease\\

$b$ & $\mu g/g$ & crowding effect of the consumer\\

$\lambda$ & $day^{-1}$ & disease transmission coefficient\\

$k_{2}$ & $g/\mu g/day$ & effect coefficient of the toxin on the consumer mortality\\

$\mu$ & $day^{-1}$ & natural mortality of consumer\\
\\
$\eta$ & $day^{-1}$ & disease related mortality of consumer\\
\\
$a_{1}$ & $L/g/day$ & uptake coefficient of resource\\

$\sigma_{1}$ & $day^{-1}$ & depuration coefficient of resource\\

$a_{2}$ & $L/g/day$ & uptake coefficient of consumer\\

$\sigma_{2}$ & $day^{-1}$ & depuration coefficient of consumer\\

$T$ & $\mu g/L$ & toxin concentration in the environment\\

\noalign{\smallskip}\hline
\end{tabular}
 \end{table}

The environmental toxicity is responsible for reducing the growth and reproduction of species in several ways. It can cause habitat degradation via changing chemical properties like salinity, acidity of marine surface and also hamper the growth of the primary producers like phytoplankton by changing the nutrient cycle \citep{cheevaporn2003water, roberts2013ocean, zeng2015positive}. Thus we consider the effect of toxin on resource's growth rate, $b(u)$, to be a monotonically decreasing function of the toxin body burden, $u$, i.e., $max (0, 1-\alpha_2 u)$ \citep{huang2013model, huang2015impact, thieme2003princeton}. So the new maximum reproduction rate is given by $\alpha_{1} b(u) = \alpha_{1} max (0, 1-\alpha_2 u)$ (see Fig. \ref{responses}a), which decreases linearly with $u$ upto the threshold value of $\displaystyle \frac{1}{\alpha_2}$, after which it becomes zero and so the resource stops growing. $\alpha_{2}$ is the effect coefficient of the toxin on the growth rate of the resource.

Furthermore, toxicants lead to decrease in the food conversion efficiency of the consumer \citep{huang2015impact, garay2013more}. We assume the consumer's reproduction efficiency to be a linearly decreasing function of the consumer body burden, $v$, given by $e(v)=\beta_1 max(0, 1-\beta_2 v)$ (see Fig. \ref{responses}C), which becomes zero after the threshold value $\displaystyle \frac{1}{\beta_2}$.  $\beta_{1}$ is the maximum conversion efficiency of the consumer and $\beta_{2}$ is the effect coefficient of the toxin on the consumer reproduction \citep{ huang2015impact}. 

Environmental toxins decreases immunity of species against diseases \citep{de1994impairment, de1996impaired, ross1996contaminant} which increases the transmission rate. Keeping this in mind, we incorporate the effect of toxin on disease transmission. This is assumed to be a monotonically increasing function of the consumer body burden, $v$, \citep{wang2018switching} and is given by $\lambda(v)= (1+ \displaystyle \frac{mv}{b+v})\lambda $ which eventually saturates to a limiting value $(1+m)\lambda$ (see Fig. \ref{responses}D). Here, the parameter $m$ is the effect coefficient of the toxin on the transmission and $b$ is the crowding effect of the consumer, the reciprocal of which corresponds to how fast the transmission rate reaches to its limiting value.

All of the resource, susceptible and infected consumer are assumed to have toxin dependent morality term in addition to their natural mortality, which are linear functions of their toxin body burdens (Fig. \ref{responses}b). The form of the mortality terms are $\mu_{1}(u)=p+k_{1}u$ and $\mu_{2}(v)=\mu +k_{2}v$ respectively, $k_{i}, (i=1,2)$ being the effect coefficients of the toxin related mortality for resource and consumers respectively. See Table \ref{par_table} for the description and units for all parameters and state variables mentioned so far.

\subsection{Non-dimensionalization and quasi-steady state approximation}

 We introduce the following dimensionless variables and parameters: 
\begin{equation*}
    \begin{split}
      \Bar{x} &= \alpha_{3}x, \quad   \Bar{y} = \frac{a \alpha_{3}}{\alpha_{1}}y, \quad \Bar{t}= \alpha_{1}t, \quad
       \Bar{u}= \alpha_{2}u, \quad \Bar{v}= \beta_{2}v,\\
     &\Bar{T}= \frac{\alpha_{2} a_{1}}{\sigma_{1}} T, \quad \Bar{k_{1}}= \frac{k_{1}}{\alpha_{1} \alpha_{2}}, \quad 
     \Bar{k_{2}}= \frac{k_{2}}{\alpha_{1} \beta_{2}}, \quad \Bar{p}= \frac{p}{\alpha_{1}}, \quad  \Bar{h}= H\alpha_{3},\\
     &\Bar{\beta_{1}}= \frac{a \beta_{1}}{\alpha_{1}}, \quad \Bar{\beta_{2}}= \frac{a \beta_{2}}{\alpha_{2} \sigma_{1}}, \quad \Bar{\sigma_{2}}= \frac{\sigma_{2}}{\sigma_{1}}, \quad \Bar{\mu}= \frac{\mu}{\alpha_{1}}, \quad
     \Bar{\eta}= \frac{\eta}{\alpha_{1}},\\
     &\Bar{\lambda}= \frac{\lambda}{\alpha_{1}}, \quad \Bar{b}= b \beta_{2}, \quad c= \frac{a_{2} \beta_{2}}{a_{1} \alpha_{2}}, \quad \epsilon= \frac{\alpha_{1}}{ \sigma_{1}}
       \end{split}
\end{equation*}







Substituting into the equations (\ref{eq1}a, d, \ref{eq2} and \ref{eq4}) and omitting the bars, we rewrite the system of equations as follows:

\begin{subequations}
\label{finalequations}
    \begin{equation}
       \frac{dx}{dt}= max(0, 1-u)\frac{x}{1+x} -(k_{1}u+ p)x     -\frac{xy}{h+x} 
    \end{equation}

    \begin{equation}
        \frac{dy}{dt}= \beta_{1} max(0, 1-v) \frac{xy}{H+x} -(k_{2}v+ \mu)y-\eta iy
    \end{equation}
    
    \begin{equation}
        \frac{di}{dt}= ((1+\frac{mv}{b+v})\lambda -\eta)i(1-i)
                       -\beta_{1} max(0, 1-v) \frac{xi}{h+x} 
    \end{equation}

    \begin{equation}
    \epsilon\frac{du}{dt}=  (T-u) -
      \epsilon max(0, 1-u)\frac{u}{1+x}
    \end{equation}

    \begin{equation}
       \epsilon \frac{dv}{dt}= cT - \sigma_{2}v + 
\frac{x}{h+x} (\beta_{2}u - \epsilon \beta_{1} v max(0, 1-v))
    \end{equation}
    
    \end{subequations}

The dynamics of the toxin body burden operates on a faster timescale compared to the species biomass growth. So the depuration rate of the toxin is higher compared to the reproduction of the resource. Our parameter $\epsilon$ is the ratio of the resource reproduction rate to the depuration rate of the toxin and so must be very small. So letting $\epsilon$ tends to zero, equation \ref{finalequations} (d, e) approaches to a quasi-steady state, which are given below:
    
    \begin{subequations}
    \begin{equation}
       u=T 
    \end{equation}
     \begin{equation}
        v=\frac{cT}{\sigma_{2}}+\frac{\beta_{2}T}{\sigma_{2}}\frac{x}{h+x}
    \end{equation}
\end{subequations}

Substituting the quasi-steady states of the body burden equations, our simplified model becomes:

\begin{subequations}
\label{eq5}
    \begin{equation}
       \frac{dx}{dt}= (\frac{max(0, 1-u)}{1+x})x-(k_{1}u+p)x-\frac{xy}{h+x} 
    \end{equation}
     \begin{equation}
        \frac{dy}{dt}= \beta_{1} max(0, 1-v) \frac{xy}{h+x}-(k_{2}v+\mu)y-\eta iy
    \end{equation}
     \begin{equation}
     \frac{di}{dt}=[(1+\frac{mv}{b+v})\lambda-\eta]i(1-i)- \beta_{1} max(0, 1-v) \frac{x}{h+x}i   
    \end{equation}
\end{subequations}

This model is further analyzed in the remaining part of the paper to study the role of environmental toxins in infectious disease dynamics in ecosystems.

\subsection{Model calibration and analysis}

For model analysis, all of the parameters are chosen from published literature \citep{huang2015impact, hilker2008disease} which includes calibrated as well as hypothetical sets of values. In the absence of disease, our model reduces to that of \cite{huang2015impact} and in toxin free environment, it is equivalent to \cite{hilker2008disease}. So the parameters related to toxin are chosen from the former while the disease related ones are chosen from the latter.

First, the mathematical proof of positive invariance and boundedness of the solutions of our model was carried out (see Appendix \ref{appendixB}). In order to explore the role of disease and toxicity on the steady state population dynamics, broad range of epidemiological parameters was chosen while keeping the ecological parameters fixed to compare the different dynamical scenarios with different level to toxicity and disease.  We perform several one and two-parameter bifurcation diagrams to track the changes in the steady state population, with varying specified parameters through the MATCONT 6p11 \citep{dhooge2008new} in MATLAB software.      

\section{Results}\label{results}

We use bifurcation analyses as a tool to address the questions outlined earlier. We also attempt to provide intuitive explanations of the different dynamical phenomena observed in the system. In the rest of the paper, $EQ_{1}$ denotes the disease-free coexistence equilibrium and $OSC_{1}$ denotes disease-free oscillations. The endemic equilibrium and population cycles are indicated as $EQ_{2}$ and $OSC_{2}$ respectively. Finally, the ecological extinction of the host is denoted by $CE_{1}$ and epidemiological extinction by $CE_{2}$. 

\subsection{Effect of toxin on disease dynamics}

\begin{figure}[h]
    \begin{center}
        \includegraphics[width=1\textwidth]
        {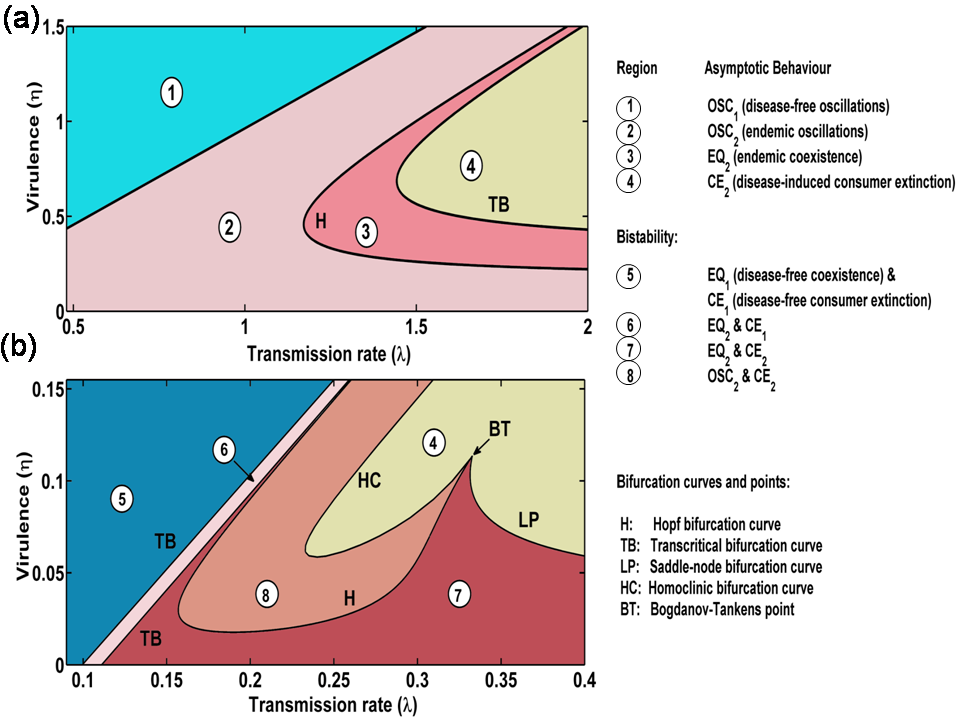}
    \end{center}
    \caption{ Two-parameter bifurcation diagram with respect to transmission rate ($\lambda$) and virulence ($\eta$) showcasing eight different dynamical regimes for environmental toxin levels (a) T=0.1, (b) T=0.1875. The other parameter values are 
$b_{1}=1,$ $ b_{2}=4, $ $h=0.6,$ $ m=0.1,$ $ b=1,$ $ c=1.5,$ $ \sigma_{2}=1, $ $k_{1}=1,$ $p=0.1,$ $ k_{2}=0.2,$ $ \mu = 0.02$.}
    \label{lameta}
\end{figure}

The process of disease progression and elimination mainly depends on the two key epidemiological parameters, namely transmission rate ($\lambda$) and the virulence ($\eta$). We compare the dynamical behaviors of our system in the $\lambda-\eta$ plane for low and high contamination levels (see Fig. \ref{lameta}). When toxin is low ($T=0.1$, Fig. \ref{lameta}A), disease is introduced into the disease-free oscillations (region \textcircled{1}) with an increase in $\lambda$ leading to endemic cycles in region \textcircled{2}. This is followed by a Hopf bifurcation (H) as a result of which the cycle stabilizes in region \textcircled{3}. Further increase in $\lambda$ leads to epidemiological extinction of the consumers in region \textcircled{4} by a transcritical bifurcation (TB). This behavior is qualitatively similar to the results demonstrated by \cite{hilker2008disease} who analyses the same model in the absence of toxicity. It is interesting to note that at higher virulence, $\eta$, disease establishes in the system only for reasonably high $\lambda$. This is because high virulence eliminates the infected host from the system resulting in eradication of the disease.

\begin{figure}[H]
    \begin{center}
        \includegraphics[width=1\textwidth]
        {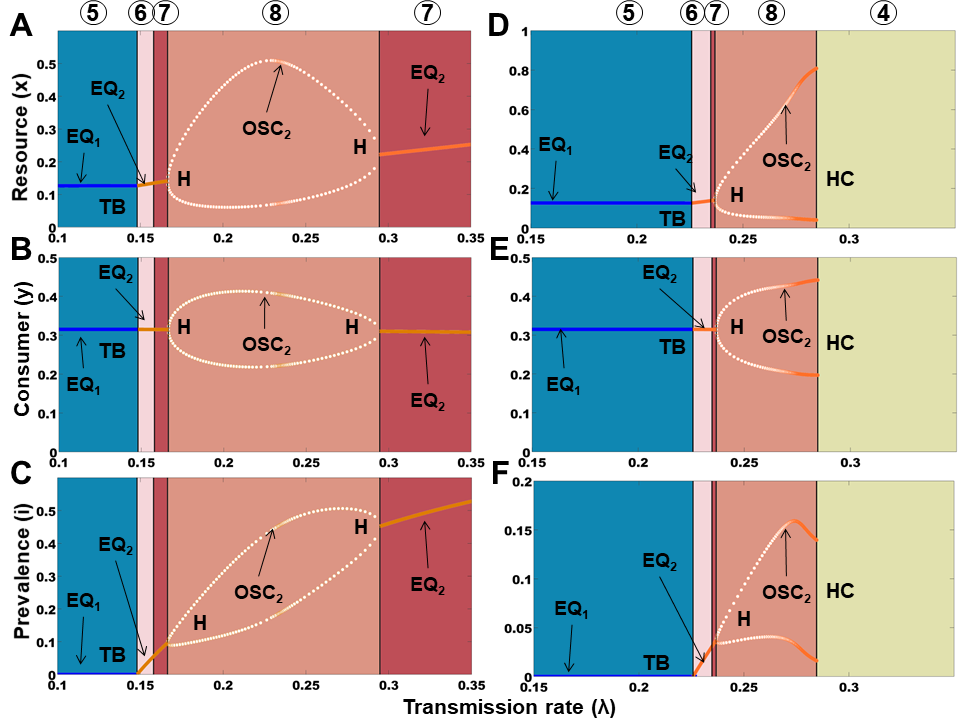}
    \end{center}
    \caption{ Effect of the transmission rate ($\lambda$) for virulence levels (A-C) $\eta=0.05$ and (D-F) $\eta=0.13$ under the high environmental toxin level ($T=0.1875$). Light blue and orange curves represent the disease-free ($EQ_{1}$) and endemic equilibrium ($EQ_{2}$) respectively. Orange circles represent the maximum and minimum population density of the endemic cycle ($OSC_{2}$). Parameter values are the same as in Fig. \ref{lameta}, and bifurcation points have their usual meaning as mentioned in Fig. \ref{lameta}.   }
    
    \label{homo}
\end{figure}

The dynamical behaviour of our system changed significantly with an increase in the toxin level (for $T=0.1875$, Fig. \ref{lameta}B). Now, when $\lambda$ is very low, the system exhibited bistability between two alternative stable states, $EQ_{1}$ and $CE_{1}$ (region \textcircled{5}). Further, moving along the $\lambda$ axis, the equilibrium $EQ_{1}$ alters its stability with $EQ_{2}$ resulting is persistence of disease in region \textcircled{6}. For low virulence ($\eta$), further increase in $\lambda$ will eventually shift the system dynamics to region \textcircled{7} via a transcritical bifurcation (TB) where the system can switch between two alternative stable states, $EQ_{2}$ and $CE_{2}$. Here,  depending on initial disease prevalence, the consumer survives with a partially infected population or will go to extinction (see \ref{app2}). On the other hand, for higher virulence ($\eta$), on moving along the $\lambda$ axis, $EQ_2$ becomes unstable and endemic oscillation start through a Hopf bifurcation. So bistability between the states $OSC_{2}$ and $CE_{2}$ is observed in region \textcircled{8}. These oscillations can either become stable leading `bubbling effect' as demonstrated in Fig. \ref{homo}A-C or their amplitude increases with increasing $\lambda$ until it collapses suddenly to region \textcircled{4} through a homoclinic bifurcation (see Fig. \ref{homo}D-F). For the sake of clarity, we do not plot the consumer extinction equilibria ($CE_1$ and $CE_2$) in this figure.

\subsection{Interplay between toxin and disease}

\begin{figure}[H]
    \begin{center}
        \includegraphics[width=1\textwidth]
        {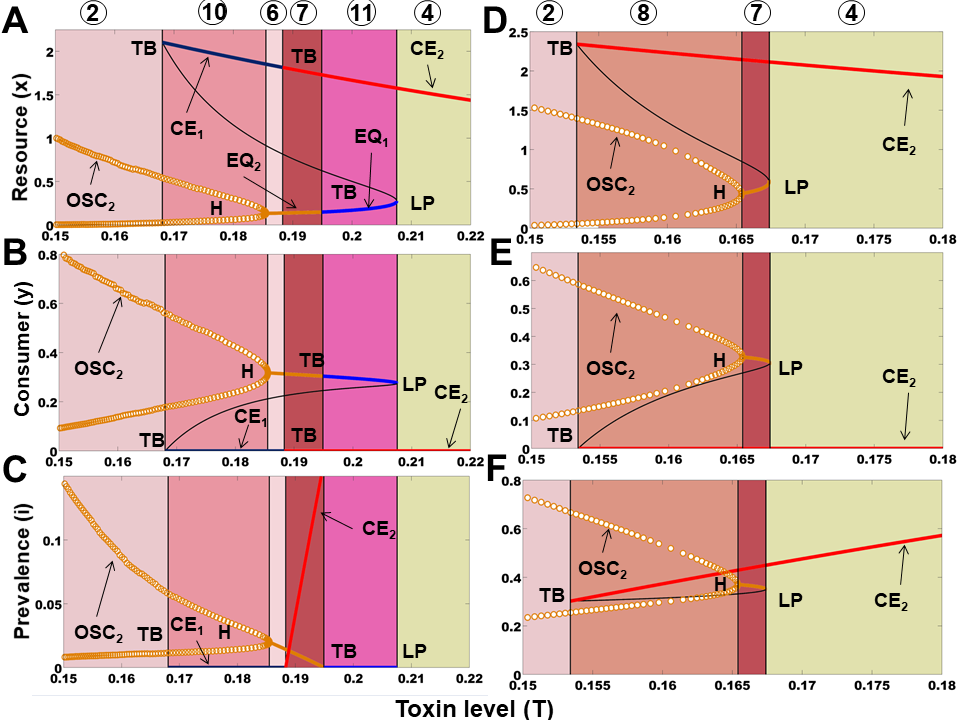}
    \end{center}
    \caption{ Effect of changing toxin concentration on the resource ($x$), the consumer ($y$), and prevalence ($i$) for different levels of disease transmission, (A-C) $\lambda=0.3$ and (D-F) $\lambda=0.5$. Parameter values are the same as Fig. \ref{lameta}. Orange and light blue curves indicate the endemic ($EQ_2$) and disease-free coexistence equilibrium ($EQ_1$) respectively. Orange circles represent the maximum and minimum population density of the endemic cycle ($OSC_2)$. Deep blue and red curves indicate the ecological and epidemiological extinction of the consumers ($CE_{1}$ and $CE_{2}$, respectively). Black curves represent the unstable equilibrium branch.           
    }
    \label{codim1}
\end{figure}

To better understand the environmental toxin's effect on the asymptotic resource and consumer dynamics, we plot one parameter bifurcation diagrams with respect to toxicity ($T$), for different transmission rates (e.g., for $\lambda=0.3$ and $0.5$) (see Fig. \ref{codim1}). When $\lambda = 0.3$ (Fig. \ref{codim1}(A-C)), the system is in endemic oscillations ($OSC_{2}$) state for low level of toxin. Increasing toxicity introduces an alternative stable disease-free consumer extinction state $CE_{1}$ (Fig. \ref{app2}) such that the consumer is unable to persist for low initial consumer population. Such toxicity induced Allee effect has been also demonstrated in the previous studies \citep{huang2015impact, banerjee2019effect}. The oscillations of the endemic cycle becomes stable ($EQ_2$) with increasing $T$ which is followed by a transcritical bifurcation leading to eradication of disease ($EQ_1$). Alternatively, the other equilibrium state $CE_1$, also undergoes transcritical bifurcation leading to disease induced consumer extinction, $CE_2$. Such dynamics leads to three more different types of bistability for changing toxin concentration as demonstrated in Fig. \ref{codim1} (A-C). The bistability between $EQ_1$ and $CE_2$ is especially noteworthy as depending on the disease prevalence, either the system exists in disease free coexistence or if the disease persists, it leads to consumer extinction (see Fig. \ref{app1}).  

Analysis reveals that on increasing toxin concentration, there may be an abrupt transition due to saddle-node bifurcation (LP), which leads to vanishing of the coexistence equilibrium thus rendering the consumer's epidemiological extinction ($CE_{2}$) as the only stable state of the system. This transition is irreversible, i.e., once the system has passed the critical threshold (LP), decreasing toxin concentration can no longer return the system to the coexistence state. Further, it is important to note that the elimination of the disease from the endemic state ($EQ_2$) due to toxin may thus be a precursor to an irreversible extinction of the consumers. For the case of high transmission rate ($\lambda=0.5$, Fig. \ref{codim1}D-F), although increased toxin still leads to disease induced consumer extinction, the route to such extinction differs. For instance, there is no disease free coexistence in this case unlike the previous one.

\begin{figure}[H]
    \begin{center}
        \includegraphics[width=1\textwidth]
        {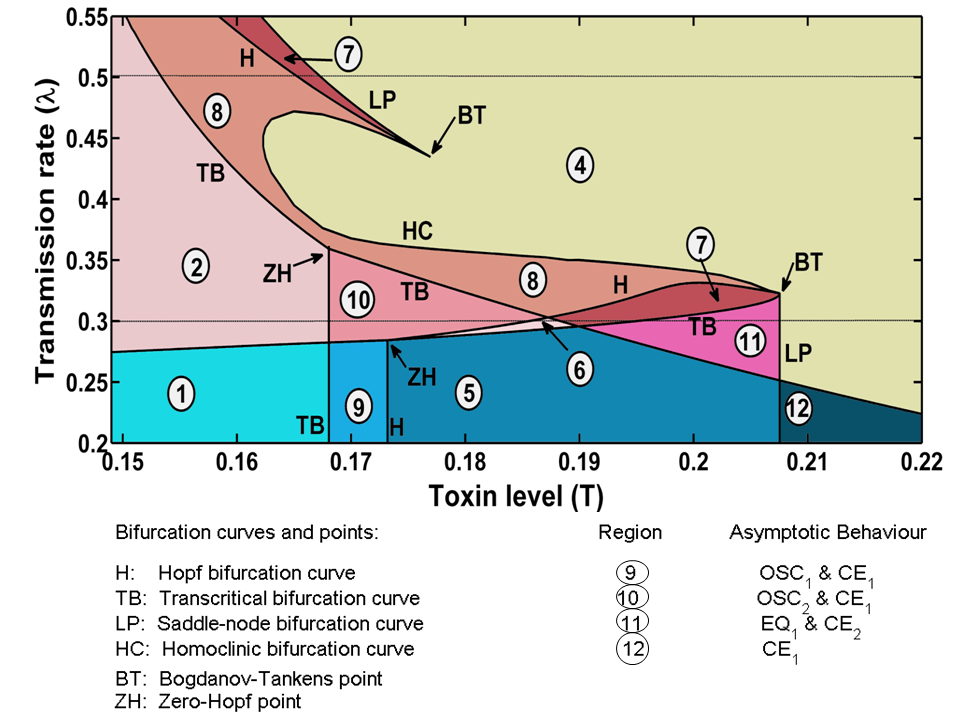}
    \end{center}
    \caption{ Two-parameter bifurcation diagram for environmental toxin level (T) and disease transmission rate ($\lambda$), demonstrating eleven different dynamical regimes. Regions \textcircled{1}-\textcircled{8} are as indicated in Fig. \ref{lameta}. Two black horizontal lines indicate the parameter values for which bifurcation diagrams in Fig. \ref{codim1} are drawn. Here $\eta=0.2$ and other parameters are same as in Fig. \ref{lameta}.}
    \label{Tlam}
\end{figure}





The results so far points out the intricate relationship between toxin and disease dynamics. In order to achieve a holistic insight into how disease and toxicity jointly shape the community structure of our system, we carry out a two-parameter bifurcation analysis in the $T - \lambda$ plane (see Fig. \ref{Tlam}). In this parametric plane, we identify all the different regions of asysmptotic behaviour  exhibited in Fig. \ref{lameta} (except \textcircled{3}). In addition, we find four new regions. Region \textcircled{9} represents toxin induced bistability between disease free population cycle, $OSC_1$ and disease free consumer extinction, $CE_1$, which occurs only when transmission rate is very low. Very high toxin level lead to loss of such bistability and $CE_1$ is the only possible state as exhibited in region \textcircled{12}. At intermediate transmission rate, disease is introduced into the population cycles making it endemic oscillations, $OSC_2$, in region \textcircled{10}. Moving along increasing toxin axis, the cycle stabilizes and becomes disease free as system exhibits alternative stable states between disease free coexistence, $EQ_1$, and disease induced consumer extinction, $CE_2$, in region \textcircled{11}. An illustration of the equilibrium dynamics in different regions along the black horizontal line has been provided in the earlier Fig. \ref{codim1}.  

Overall, it is observed that for lower values of transmission rate, $\lambda$, the system is in disease-free states represented by different shades of blue in Fig. \ref{Tlam}. If we increase the toxin level, the system moves through the regions of different asymptotic behaviour to eventually consumer extinction. When transmission rate ($\lambda$) is high, the disease induced consumer extinction (region \textcircled{4}) occurs for much less toxin level, $T$.

\section{  Discussion }
\label{discussion}

Disease and chemical pollution are important concerns for many ecosystems worldwide \citep{lafferty2003should, lafferty2004diseases, van2009emerging} but studies integrating the two are rare. We used the approach prescribed by \cite{huang2013model,huang2015impact} to model the combined impact of environmental toxin and disease on consumer dynamics. We carried out bifurcation analysis, mainly with respect to environmental toxin parameter and disease related parameters to study the model behaviour. The results showcased in our study demonstrate the emergence of different dynamical scenarios and may help to contribute to our understanding of community ecology in the face of current anthropogenic changes. 

To understand the impact of toxin on disease dynamics, we compare the behaviour of the system in the transmission-virulence plane for different toxin levels (see Fig. \ref{lameta}). With increasing toxin level, the disease may be established in the system for relatively lower transmission rate. This is expected because increasing environmental toxin increases disease transmission rate in our model. Furthermore, under low toxin concentration, when the system exhibits disease induced consumer extinction, it is the only stable state. On the contrary, when toxin level is high, if the initial disease prevalence is low, disease induced extinction may be avoided and the system may end up in either endemic coexistence equilibrium or population cycles depending on the transmission rate (see Fig. \ref{app2}.B, C). The fact that disease causes this destabilization from equilibrium to cycles is noteworthy. This is because it is contrary to the stabilizing effect of disease as demonstrated in not only earlier studies \citep{hilker2008disease} but also when toxicity is low. The cycles collapse under high transmission rendering disease induced consumer extinction as the only possible stable state. Related interesting observation is that when virulence ($\eta$) is low, these cycles again stabilize on increasing transmission thus producing bubbling effect (see Fig. \ref{homo}A-C).  


A better understanding of the interaction between disease and toxin is achieved on studying the one parameter bifurcation diagram with respect to toxin and the system behaviour in the transmission-toxin parameter plane (Fig. \ref{Tlam}). While increasing toxin amplifies the risk of disease as has been noted in the earlier paragraph, toxicity also eradicate the prevalent disease from the system. This is observed in region \textcircled{11} where one of the bistable states, the endemic equilibrium, changes to disease free coexistence under increasing toxin (Fig.\ref{codim1}.A, \ref{Tlam} ). Here, the consumer asymptotically may either become disease-free or extinct, depending on the initial prevalence level (Fig. \ref{app1}.C). Abrupt disease-induced consumer extinction may also be observed. The toxin level at which such extinction occurs is higher when transmission is low which can be attributed to the synergistic effect of toxin on transmission in our model. Population oscillation and interestingly even disease-free coexistence can be precursor of such abrupt host extinction in the presence of toxin.  
      

Toxicity introduces disease-free host extinction as an alternative stable state to the population cycles (Fig. \ref{codim1}.A-C). For instance, one can note the transition from $OSC_1$ to region \textcircled{9} where both $OSC_1$ and $CE_1$ can exist and be stable (Fig. \ref{app1}.B). Similarity in this behaviour to that observed by \cite{huang2015impact} is because of the fact that in the absence of disease our model reduces to that of the earlier one. In the presence of disease, this is translated to an interesting property whereby the prevalence of disease helps the host survive which would otherwise become extinct. However, for very high transmission, this is no more completely true as then the survival would depend on the initial prevalence of disease. Too high disease prevalence could then lead to disease-induced host extinction (see Fig. \ref{Tlam}).



Overall our study throws light into the role of toxin in initiation and progression of an epidemic. Although the results presented here are significant in the context of disease dynamics, two key limitations of our model must be noted which could be addressed in future works. The effects of the toxin on disease induced host death rate were not considered in our analysis. Additionally, the infected population may recover from the disease, which is not considered in our system. Our work highlights the need to undertake more studies which will help comprehend the interaction between anthropogenic changes and disease in ecosystems and the non-linearity therein. Although management recommendation is not our direct aim but better fundamental understanding will pave the way for empirical validation and definitive actions.

\bmhead{Acknowledgments}

Arnab and Amit acknowledges Senior Research Fellowship (file no: 09/093(0190)/2019-EMR-I and 09/093(0189)/2019-EMR-I) from Council of
Scientific and Industrial Research (CSIR), India. Swarnendu was supported by the Visiting Scientist fellowship at Indian Statistical Institute, Kolkata during a part of this work. Swarnendu would also like to acknowledge his present funding under the Marie Skłodowska–Curie grant
agreement 101025056 for the project ‘SpatialSAVE’.

\bmhead{Author contribution}
Arnab and Swarnendu conceived the idea; Arnab, Swarnendu, Sabyasachi refined it; Arnab and Swarnendu led the study and designed the simulations. Arnab and Amit programmed and ran the simulations. Arnab wrote
the first draft of the manuscript. Swarnendu and Sabyasachi reviewed and edited the manuscript. All authors read and approved the final manuscript.

\section*{Declarations}
\bmhead{Conflict of interest} The authors declare no competing interests.

\begin{appendices}

\section{Non-dimensionalisation and steady-state approximation}
\label{appendixA}

We first derive the equations of the disease prevalence and toxin body burdens of both the resource and consumers respectively as follows:

\begin{equation*}
\begin{split}
\frac{di}{dt}& 
= \frac{y\frac{dI}{dt}-I\frac{dy}{dt}}{y^2}\\
& =\frac{1}{y}\frac{dI}{dt}-\frac{I}{y}(\frac{y'}{y})\\
& =\frac{1}{y}[(1+\frac{mv}{b+v})
\frac{\lambda SI}{y}-(\mu+k_{2}v)I-\eta I]-
i[e(v)\frac{ax}{H+x}-(\mu+k_{2}v)-\eta \frac{I}{y} ]\\
& =[1+\frac{mv}{b+v}]\frac{\lambda Si}{y}-
(\mu+k_{2}v)i-\eta i -
i[e(v)\frac{ax}{H+x}-(\mu+k_{2}v)-\eta i ]\\
& =[1+\frac{mv}{b+v}]\frac{\lambda i(1-i)y}{y}
-e(v)\frac{ax}{H+x} i -\eta i(1-i)\\
& =[1+\frac{mv}{b+v}]\lambda i(1-i)-
e(v)\frac{ax}{H+x} i -\eta i(1-i)  
\end{split}
\end{equation*}

\begin{equation*}
\begin{split}
\frac{du}{dt}&
=\frac{d}{dt} (\frac{U}{x})\\
&=\frac{1}{x} \frac{dU}{dt} - 
\frac{U}{x} (\frac{1}{x} \frac{dx}{dt})\\
&=\frac{1}{x} (a_{1}Tx-\sigma_{1}U- 
\mu(u)U- \frac{axy}{H+x} u)-
u(\beta(x,u)- \mu (u)- \frac{ay}{H+x})\\
&=a_{1}T- \sigma_{1} u -\beta(x,u) u  
\end{split}
\end{equation*}

\begin{equation*}
\begin{split}
\frac{dv}{dt}&
= \frac{d}{dt} (\frac{V}{y})\\
&=\frac{1}{y} \frac{dV}{dt} - 
\frac{V}{y} (\frac{1}{y} \frac{dy}{dt})\\
&=\frac{1}{y} (a_{2}Ty-\sigma_{2}V+
\frac{axy}{H+x} u - (\mu +k_{2}v)V - \eta \frac{I}{y} V)-
v(e(v)\frac{ax}{H+x} - (\mu+k_{2}v)-\eta \frac{I}{y})\\
&=a_{2}T-\sigma_{2}v+ \frac{ax}{H+x} (u-v e(v)) 
\end{split}
\end{equation*}

\section{Proof of positivity and boundedness}
\label{appendixB}

The right hand side of the system (6 a-c) is continuously differentiable and locally Lipschitz in the first quadrant which implies the existence and uniqueness of
solutions for the system in $R_{+}^{3}$. For positive invariance we rewrite the system as:
\begin{equation}
    \frac{dX}{d\tau}=F(X)
\end{equation}
Where $X=[x,y,z]^T \in R_{+}^{3}$ and $F(X)=[F_{1}(X),F_{2}(X),F_{3}(X)]^{T}$. The solutions of the system remain in the first quadrant for any non-negative initial condition for all $\tau\geq 0$, since 
${F_{i}(X)}\mid _{X_{i}=0}\geq 0$, for all $ X_{i} =0$ where $i=1,2,3$.

To proof positivity of the solutions of the system 6 a-c,  we assume,
\begin{equation*}
Z(t)=x+y+i
\end{equation*}
Differentiating $Z$ with respect to $t$ we get,
\begin{equation*}
\begin{split}
 \frac{dZ}{dt} &
 =\frac{dx}{dt}+\frac{dy}{dt}+\frac{di}{dt}\\
 & =(\frac{[1-u]_{+}x}{1+x})-(k_{1}u+p)x-\frac{xy}{h+x}+\beta_{1}\frac{[1-v]_{+}xy}{h+x}-(k_{2}v+\mu)y\\
 & -\eta iy+[(1+\frac{mv}{b+v})\lambda-\eta]i(1-i)-\beta_{1}[1-v]_{+}\frac{x}{h+x}i 
\end{split}
\end{equation*}
Since $(\frac{[1-u]_{+}x}{1+x})< 1$, $\beta_{1}\frac{[1-v]_{+}xy}{h+x}-\frac{xy}{h+x}\leq 0$  we can write,
\begin{equation*}
    \frac{dZ}{dt}< 1-(k_{1}u+p)x-(k_{2}v+\mu)y
 -\eta iy+(1+\frac{mv}{b+v})\lambda i(1-i)-\eta i(1-i)-\beta_{1}[1-v]_{+}\frac{x}{h+x}i 
\end{equation*}
As $\frac{v}{b+v} < 1$, $k_{1}ux\geq0$,  $k_{2}vy\geq 0$, $\eta i(1-i)\geq 0$ and $\beta_{1}[1-v]_{+}\frac{x}{h+x}i\geq 0$, 
\begin{equation*}
   \frac{dZ}{dt} \leq  1-px-\mu y+(1+m)\lambda i(1-i)
\end{equation*}
For an arbitrary positive real number $N$ we get,
\begin{equation*}
    \frac{dZ}{dt}+NZ < 1-x(p-N)-y(\mu-N)+i((\lambda+m\lambda +N)-i(1+m)\lambda)
\end{equation*}
Let $N\leq min(p,\mu)$ and the maximum value of $i((\lambda+m\lambda +N)-i(1+m)\lambda)$ is $\frac{(\lambda+m\lambda+N)^2}{4(1+m)}$, 
\begin{equation*}
    \frac{dZ}{dt}+NZ < 1+\frac{(\lambda+m\lambda+N)^2}{4(1+m)}
\end{equation*}
Substituting $A=1+\frac{(\lambda+m\lambda+N)^2}{4(1+m)}$ we get,
\begin{equation*}
    \frac{dZ}{dt}+NZ < A
\end{equation*}
By diffrential inequality
\begin{equation*}
    0< Z(x,y,z) < \frac{A(1-exp(-Nt))}{N}+Z(x(0),y(0),z(0))exp(-Nt)
\end{equation*}
So for large values of t we have $0 \leq Z\leq \frac{A}{N}$. Hence the solution of the system are bounded in the positive quadrant.

\section{Bistablity in the consumer-resource system}
\label{appendixC}

\begin{figure}[h!]
    \begin{center}
        \includegraphics[width=0.9\textwidth]{ 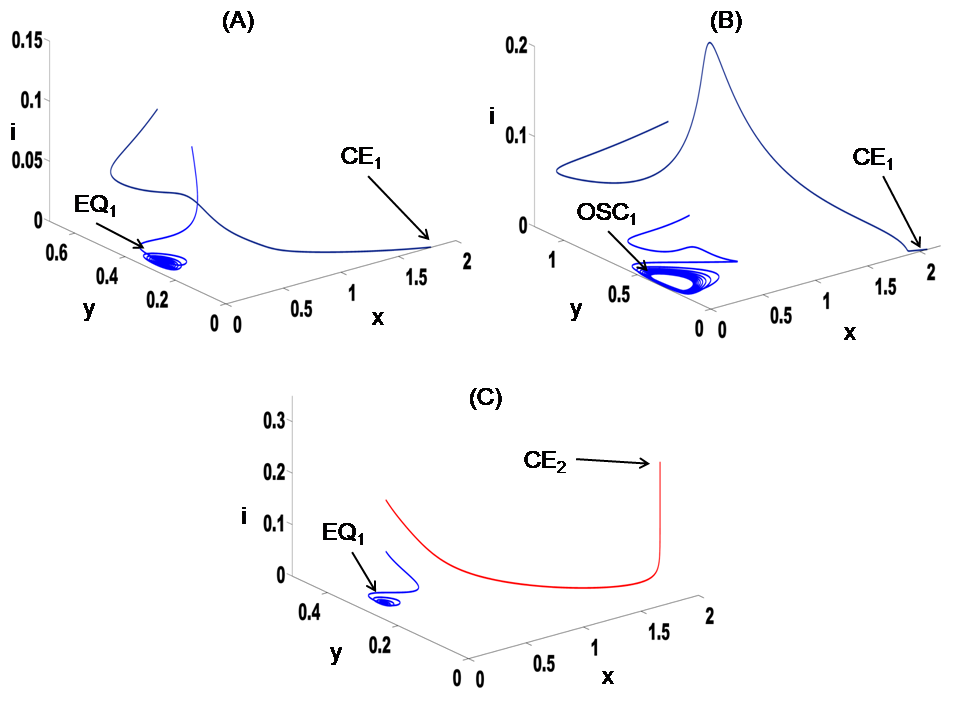}
    \end{center}
    \caption{ The solution trajectories for different bistability regions. (A) region \textcircled{5}; (parameters: $T=0.19$, $\lambda=0.25$), (B) region \textcircled{9}; (parameters: $T=0.17$, $\lambda=0.25$), (C) region \textcircled{11}; (parameters: $T=0.2$, $\lambda=0.3$). Bistability between disease-free coexistence and consumer extinction (disease-free or disease-induced) is seen.}
    \label{app1}
\end{figure}

\begin{figure}[H]
    \begin{center}
        \includegraphics[width=0.9\textwidth]{ 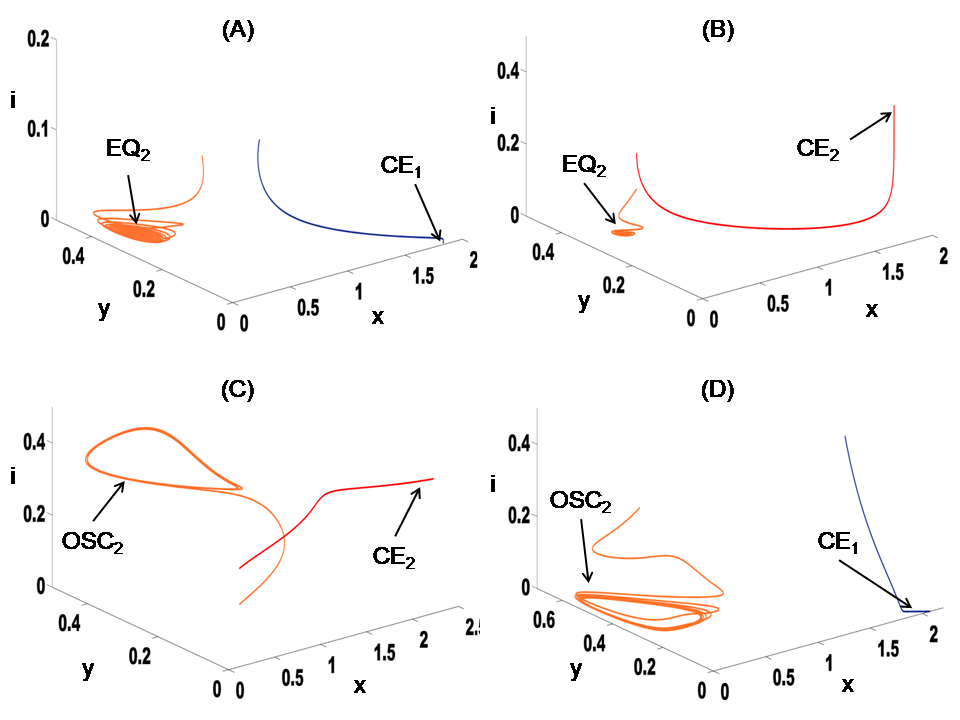}
    \end{center}
    \caption{ The solution trajectories for different bistability regions. (A) region \textcircled{6}; (parameters: $T=0.187$, $\lambda=0.3$), (B) region \textcircled{7}; (parameters: $T=0.2$, $\lambda=0.32$), (C) region \textcircled{8}; (parameters: $T=0.16$, $\lambda=0.5$), (C) region \textcircled{10}; (parameters: $T=0.17$, $\lambda=0.3$). Bistability between endemic coexistence and consumer extinction (disease-free or disease-induced) is seen.}
    \label{app2}
\end{figure}

The system exhibits bistable dynamics for various parameter regimes (region \textcircled{5}-\textcircled{11}).  Fig. \ref{app1} demonstrates the bi-stable regimes for low level of disease transmission rate, whereas Fig. \ref{app2} for high level of transmission rate.  

\newpage
\end{appendices}


\bibliography{sn-bibliography}


\end{document}